\documentclass[twocolumn,showpacs,superscriptaddress,amsmath,amssymb,prl,10pt,showkeys,aps]{revtex4-1}
\usepackage{graphicx}
\usepackage{dcolumn}
\usepackage{bm}
\usepackage{hyperref}
\usepackage{amsthm}

\usepackage[normalem]{ulem}
\usepackage[usenames,dvipsnames]{color}

\makeatletter
\newcommand*{\addSI}{%
  \close@column@grid
  \cleardoublepage
  \twocolumngrid
}
\makeatother

\def\eqref#1{(\ref{#1})}
\def\angstrom{{\mbox{\AA}}}
\newcommand{\un}[1]{\,\mathrm{#1}}

\newtheorem*{theorem*}{Theorem}
\newtheorem*{corollary*}{Corollary}

\definecolor{tangerine}{rgb}{0.944,0.522,0}
\definecolor{verde}{rgb}{0.,0.6,0}
\definecolor{rosso}{rgb}{0.9,0.0,0.2}
\definecolor{magenta}{rgb}{0.9,0.2,0.9}

\newcommand{\editor}[2]{%
  \expandafter\newcommand\csname #1note\endcsname[1]{%
    \textcolor{#2}{(\textbf{#1:} ##1)}}%
  \expandafter\newcommand\csname #1\endcsname[1]{%
    \textcolor{#2}{##1}}%
  \expandafter\newcommand\csname #1cancel\endcsname[1]{%
    \textcolor{#2}{\sout{##1}}}%
  \expandafter\newcommand\csname #1change\endcsname[2]{%
    \textcolor{#2}{\sout{##1} ##2}}%
  \newenvironment{#1text}{\color{#2}}{\color{black}}
}

\editor{SB}{tangerine}
\editor{RB}{blue}
\editor{LE}{verde}
\editor{FG}{rosso}

\begin{document}

\title{Theory and numerical simulation of heat transport in multi-component systems}

\author{Riccardo Bertossa}
\affiliation{SISSA -- Scuola Internazionale Superiore di Studi Avanzati, Via Bonomea 265, 34136 Trieste, Italy}
\author{Federico Grasselli}
\affiliation{SISSA -- Scuola Internazionale Superiore di Studi Avanzati, Via Bonomea 265, 34136 Trieste, Italy}
\author{Loris Ercole}
\altaffiliation[Present address: ]{Theory and Simulation of Materials (THEOS), and National Centre for Computational Design and Discovery of Novel Materials (MARVEL), \'Ecole Polytechnique F\'ed\'erale de Lausanne, CH-1015 Lausanne, Switzerland.}
\affiliation{SISSA -- Scuola Internazionale Superiore di Studi Avanzati, Via Bonomea 265, 34136 Trieste, Italy}
\author{Stefano Baroni}\email{baroni@sissa.it}
\affiliation{SISSA -- Scuola Internazionale Superiore di Studi Avanzati, Via Bonomea 265, 34136 Trieste, Italy}
\affiliation{CNR -- Istituto Officina dei Materiali, SISSA, 34136 Trieste}

\date{\today}

\begin{abstract}
  The thermal conductivity of classical multi-component fluids is seemingly affected by the intrinsic arbitrariness in the definition of the atomic energies and it is ill-conditioned numerically, when evaluated from the Green-Kubo theory of linear response. To cope with these two problems we introduce two new concepts: a \emph{convective invariance} principle for transport coefficients, in the first case, and \emph{multi-variate cepstral analysis}, in the second. A combination of these two concepts allows one to substantially reduce the noise affecting the estimate of the thermal conductivity from equilibrium molecular dynamics, even for one-component systems.
\end{abstract}



\maketitle
The transport properties of macroscopic systems are determined by the dynamics of \emph{conserved currents}, \emph{i.e.} of the long-wavelength components of the currents associated to the densities of conserved extensive variables \cite{Kadanoff1963,*Foster1975,Baroni2018}. Let $J^i$ be the macroscopic average of the $i$-th conserved current, which from now on we dub a (conserved) \emph{flux}. 
In the case of heat transport in an $M$-component fluid, the relevant conserved quantities are the total energy and the total number (or mass) of molecules of each independent component. As the total-mass flux is the total momentum, which is also a constant of motion, the number of relevant conserved fluxes is reduced from $M+1$ to $M$: energy, which we label as the zero-th, and $M-1$ \emph{convective} fluxes, which can be identified with \textit{any} independent linear combinations of the molecular mass or number fluxes. In the linear regime, the relevant fluxes are linear combinations of the corresponding conjugate \emph{affinities}, $F^i$, defined as the gradients of the intensive variables conjugate to the conserved ones. These are the inverse temperature for the energy and the chemical potential divided by the temperature for the molecular numbers. The resulting Onsager relations \cite{Onsager1931a,*Onsager1931b} read:
\begin{equation}
J^i = \sum_{j=0}^{M-1} \Lambda^{ij} F^j, \label{eq:Onsager}
\end{equation}
where, in order to simplify the notation, the vector character of fluxes and affinities has been overlooked or their Cartesian indices are incorporated in the suffixes. The Green-Kubo (GK) theory of linear response {\cite{Green1952,*Green1954,Kubo1957a,*Kubo1957b,Baroni2018}} states that the $\Lambda$ matrix in Eq.~\eqref{eq:Onsager} can be expressed in terms of the time correlation functions of the various \textit{flux processes}, $\mathcal{J}^i$, which are phase-space functions, as:
\begin{equation}
\Lambda^{ij} = \frac{V}{k_B} \int_0^\infty \left\langle \mathcal{J}^i(t) \mathcal{J}^j(0) \right \rangle dt, \label{eq:L_def}
\end{equation}
where $k_B$ is the Boltzmann constant, $V$ is the system's volume, $\langle\cdot\rangle$ indicates a canonical average, and where the system is in thermodynamic \textit{equilibrium} \cite{Baroni2018}. Here and in the following, processes and their samples will be denoted by by calligraphic letters, as in ``${\mathcal X} $'', while their average properties by roman ones, as in ``$ X $''. For instance, $J^i = \langle \mathcal{J}^i \rangle$, which vanishes in the absence of perturbations.
The heat conductivity is defined as the ratio between the heat flux, $J^Q$, and the temperature gradient that generates it: $J^Q=-\kappa\nabla T$, in the absence of net particle flow. The heat flux is defined as $\mathcal{J}^Q= \mathcal{J}^0-\sum_{s=1}^{M}h^s \mathcal{J}^s$, where $\mathcal{J}^0$ and $\mathcal{J}^s$ are the energy and particle-number flux samples of all molecular species, respectively, and $h^s$ are the corresponding partial enthalpies \cite{DeGroot1984}. 
{The energy flux is defined in terms of atomic positions, $\mathbf{R}_n$, velocities, $\mathbf{V}_n$, energies $\epsilon_n$ as \cite{Irving1950,Baroni2018}:
\begin{equation}
    \mathcal{J}^0 = \frac{1}{V} \left[ \sum_n \mathbf{V}_n \epsilon_n 
    + \sum_{n,m} (\mathbf{R}_n - \mathbf{R}_{m}) \mathbf{F}_{nm}\cdot \mathbf{V}_{n} \right] \label{eq:eflux}
\end{equation}
where $\mathbf{F}_{nm} = -\partial \epsilon_{m} / \partial \mathbf{R}_n$ and $n$ runs over all the atoms.}

In solids and one-component fluids, energy is the only conserved quantity relevant to heat transport because the convective fluxes either vanish or do not contribute to energy transport \cite{Baroni2018}. In these cases, the thermal conductivity is basically given by $\kappa=\Lambda^{00}/T^2$. As the energy flux{, Eq.~\eqref{eq:eflux},} is obtained via the continuity equation from the energy density, which is ill-defined at the atomic scale, it has long been feared that no quantum-mechanical expression for the heat conductivity could be obtained from first principles. This apparent conundrum was solved only recently by the introduction of a \emph{gauge invariance} principle for transport coefficients, according to which different energy densities integrating to the same total energy give rise to fluxes that differ one from the other by the total derivative of a bounded vector, which does not contribute to the value of $\Lambda^{00}$ in Eq.~\eqref{eq:L_def} \cite{Marcolongo2016,*Ercole2016,Baroni2018}. The situation is not nearly as clear when one considers the intrinsic indeterminacy in the definition of the atomic energies. When the energies of all the atoms of a same species, say the $s$-th, are shifted by a same \emph{self-energy}, $\delta^s$, not depending on the atomic environment (\textit{i.e.} $\epsilon_n \to \epsilon_n + {\delta^{s(n)}}$, where $n$ is an atomic index), it is to be expected that all the transport properties remain unchanged. For instance, in a quantum-mechanical simulation, the heat conductivity cannot depend on whether atomic cores contribute to the definition of the atomic energy, as they would in an all-electron calculation, or not, as they would when using pseudo-potentials. In the latter case, the energy of isolated atoms would depend on the specific form of pseudo-potential adopted, which is to a large extent arbitrary, but the heat conductivity in all cases should not. When the atomic energy of the $s$-th species is shifted as above, the energy flux is modified as:
$ \mathcal J^0\to \mathcal J^0 +  {\sum_s \delta^s \mathcal J^s}$. When only one atomic component is present,  {$\mathcal J^s$ is proportional to the 
} total, conserved, momentum, which can be assumed to vanish identically,  {so that} $\Lambda^{00}$ is independent of atomic self-energies. A slight generalization of this argument allows one to arrive at the same conclusion in the one-component molecular case \cite{Baroni2018,Marcolongo2019}.

In multi-component fluids the  {convective fluxes of individual species are not identically vanishing.
As a consequence, the expressions for $\mathcal{J}^0$ and $\mathcal{J}^Q$ do not coincide and the Onsager coefficient $\Lambda^{0i}$ are affected by the spurious atomic self-energies, as described before. The Green-Kubo integrals $\Lambda^{Qi}$ (see Eq.~\eqref{eq:L_def}) computed from the heat flux, instead, would remain unaffected because the self-energy contributions to the partial enthalpies cancel their contribution to the energy flux. It seems therefore that spurious self-energy effects can be disposed of, at the price however of computing partial enthalpies, a rather cumbersome task \cite{Debenedetti1987,*Abteilung1987,*Sindzingre1989}, which had better be avoided.
In practice, heat conductivities are usually measured in conditions of vanishing mass transport, where the average $J^Q$ and $J^0$ do coincide, making one hope that the computation of partial quantities (energies or enthalpies, see Ref.~\citenum{Trimble1971}) can indeed be bypassed. To see this more formally, we impose that the convective fluxes in Eq.~\eqref{eq:Onsager} vanish, and solve for the energy flux. This is best achieved by partitioning the $\Lambda$ matrix into a $1\times 1$ energy block and an $(M-1)\times (M-1)$ convective block and by performing a block inversion. The resulting expression for the heat conductivity is:}
\begin{equation}
\kappa = \frac{\bar{\Lambda}^0}{T^2}, \label{eq:multi_kappa}
\end{equation}
where $\bar{\Lambda}^0=1/\bigl (\Lambda^{-1} \bigr )^{00} = \Lambda^{00}-\sum_{i,j=1}^{M-1}\Lambda^{0i} \bigl ( \Lambda^{-1}_{M-1}\bigr )^{ij}\Lambda^{j0}$ is the  {inverse of the energy block of $\Lambda^{-1}$, that is to say the} \emph{Schur complement} of the  {convective} block (SCCB) in $\Lambda$ \cite{Crabtree1969}. Using standard matrix manipulations and the bilinearity of $\Lambda^{ij}$ with respect to the fluxes, it is straightforward to verify that the SCCB, and hence the heat conductivity, is invariant with respect to the addition of \textit{any} linear combinations of  {convective} fluxes to the energy flux:
$\mathcal J^0 \to \mathcal J^0 + \sum_{m=1}^{M-1} c^m \mathcal J^m$,  {while the whole $\Lambda$ matrix is not}. We dub this remarkable property the \emph{convective invariance} of heat conductivity in multi-component systems. An important consequence of convective invariance is the independence of the  heat conductivity on atomic self-energies, thus solving the first of our problems.  {Another important consequence is that the heat conductivities computed from the heat or energy fluxes coincide, thus dispensing us from the task of computing partial enthalpies to evaluate the former.}

Having thus cleared the way, we now move to evaluating Eq.~\eqref{eq:multi_kappa} from equilibrium molecular dynamics (EMD). In order to be specific and streamline the discussion, we specialize Eq.~\eqref{eq:multi_kappa} to the two-component case:
$\kappa = \left . \left ( \Lambda^{00} - \bigl (\Lambda^{10} \bigr )^2 \middle / \Lambda^{11} \right )\middle /T^2 \right . $. This expression is very sensitive to the statistical errors affecting the matrix elements appearing therein, because it is the difference of two positive numbers whose magnitude may be comparable, and because the errors affecting each of them may be large and difficult to estimate \cite{Galamba2007,Ohtori2009b,Salanne2011,Bonella2017}.

In order to cope with the latter problem, we have  {generalized to multivariate processes our \emph{cepstral analysis} approach to evaluating transport coefficients from EMD \cite{Ercole2017,Baroni2018}.} Cepstral analysis \cite{Bogert1963,*Childers1977} is a technique, commonly used in signal analysis and speech recognition, to process the power spectrum of a time series, leveraging its smoothness and the statistical properties of its samples.   {In the one-component case, according to Eqs.~(\ref{eq:Onsager}-\ref{eq:L_def}) a transport coefficient is proportional to the zero-frequency value of the power spectrum of the appropriate flux: $\kappa\propto S(\omega=0)$, where $S(\omega)=\int_{-\infty}^\infty \mathrm{e}^{i\omega t} C(t)dt$, and $C(t) = \langle {\mathcal J}(t){\mathcal J}(0)\rangle$ is the flux time auto-correlation function. 
The Wiener-Kintchnine theorem \cite{Wiener1930,*Khintchine1934} states that $ S(\omega)$ is asymptotically proportional to the expectation of the squared modulus of the truncated Fourier transform of the flux sample: $ S(\omega) = \lim_{\tau\to\infty} \langle {\mathcal S}_\tau(\omega) \rangle$, where ${\mathcal S}_\tau(\omega) = \frac{1}{\tau}  | \tilde {\mathcal J}_\tau(\omega)|^2 $ is the \emph{sample spectrum} 
and $\tilde{\mathcal J}_\tau(\omega) = \int_0^\tau \mathcal{J}(t) \mathrm{e}^{i\omega t}dt$.
In the long-time limit, ${\mathcal S}_\tau(\omega)$ is a  process whose values are independent for $\omega\ne\omega'$ and individually distributed as ${\mathcal S}_\tau(\omega)=S(\omega)\xi(\omega)$, where $\xi(\omega) \sim \frac{1}{2}\chi^2_2$, $\chi^2_2$ being a chi-square variate with two degrees of freedom. The multiplicative nature of the noise affecting the sample spectrum suggests that the power of the noise can be reduced by applying a low-pass filter to its logarithm. 
In cepstral analysis this idea is leveraged to devise a consistent and asymptotically unbiased estimator for the the zero-frequency value of the flux power spectrum, which is proportional to the transport coefficient we are after \cite{Ercole2017,Baroni2018}.}

In the multi-component case, an EMD simulation samples $M$ stationary stochastic processes, $\mathcal J^i~(i=0,\cdots M-1)$, one for each conserved flux, which can be thought of as different components of a same multivariate process. For such a process it is customary to define a cross time-correlation function,  { {$C^{ij}(t)=\langle {\mathcal J}^i(t){\mathcal J}^j(0)\rangle$}, and a \emph{cross power spectrum}, $S^{ij}(\omega) = \lim_{\tau\to\infty} \langle {\mathcal S}^{ij}_\tau(\omega) \rangle$, where ${\mathcal S}^{ij}_\tau(\omega) = \frac{1}{\tau} \tilde {\mathcal J}^i_\tau(\omega)^* \tilde {\mathcal J}^j_\tau(\omega) $ is the \emph{sample cross-spectrum} of the multivariate process}. The Onsager coefficients of Eq.~\eqref{eq:L_def} are proportional to the zero-frequency values of the cross-spectrum, $S^{ij}_0=S^{ij}(\omega=0)$: $\Lambda^{ij}=\frac{V}{2k_B}S^{ij}_0$. Eq.~\eqref{eq:multi_kappa} shows that in order to evaluate the heat conductivity in the multi-component case one needs an efficient estimator for the  {SCCB in $S_0$. In analogy with Eq.~\eqref{eq:multi_kappa}, we indicate the SCCB in $S(\omega)$ and in $\mathcal{S}(\omega)$ as $\bar{S}^0(\omega)$ and $\bar{\mathcal{S}}^0(\omega)$, respectively, which in the following will be dubbed the \emph{reduced (sample) spectrum}. Their zero-frequency values will be labeled as $\bar{S}^0_0$ and $\bar{\mathcal{S}}^0_0$, respectively. Mind the difference between the power spectrum, denoted by $S(\omega)$, and its sample, denoted by $\mathcal{S}(\omega)$.} 

 {In analogy to the uni-variate case, the reduced sample spectrum can be shown to be a process distributed like
\begin{equation}
  \bar{\mathcal S}^{0}(\omega) \sim \bar{S}^{0}(\omega) \, \xi(\omega),  \label{eq:mean-multi-periodogram}
\end{equation} where} 
the $\xi$'s are independent identically distributed random variables, $\xi \sim \frac{1}{\nu} \chi^2_{\nu}$, $\chi^2_\nu$ being the chi-square distribution with  {$\nu=2(\ell-M+1)$ degrees of freedom, where $M$ is the number of conserved fluxes and $\ell\ge M$ is the number of flux samples used to sample the spectrum. This means that the above formulas apply to an effective sample spectrum, defined as the average over $\ell$ independent samples of the spectrum. For instance, in an isotropic fluid one has one equivalent flux process per Cartesian component ($\ell=3$); multiple flux samples can be created by either running multiple EMD trajectories for different initial conditions or, equivalently, by breaking a long trajectory into multiple segments. Eq.~\eqref{eq:mean-multi-periodogram} shows that the reduced sample spectrum is an unbiased estimator of the reduced spectrum ($\langle\xi(\omega)\rangle=1$). Unfortunately, this estimator is not consistent, in that its variance does not vanish when the length of the time series grows large, and is actually independent of it. In order to reduce the power of the noise and obtain a consistent estimator of the reduced spectrum, we apply a low-pass filter to its logarithm. To this end, one first performs a (inverse) Fourier transform of the logarithm of the reduced spectrum, and one retains a number of coefficients, $P^*$, equal to the estimated number of non-vanishing Fourier coefficients of the logarithm of the reduced spectrum. By doing so, the estimator of the heat conductivity we are after, $\mathcal{K}$, and its statistical uncertainty can be expressed as:}
\begin{equation}
    \begin{gathered}
       \mathcal{K} = \frac{V}{2k_BT^2}\mathrm{exp}\left [ \mathcal{C}_0 + 2\sum_{n=1}^{P^*-1}\mathcal{C}_{n} -L_0\right ] \\
        \frac{\Delta\kappa}{\kappa} = \sigma_0 \sqrt{\frac{4P^*-2}{N}},
    \end{gathered} \label{eq:nutshell}
\end{equation}
 {where the $\mathcal{C}$s are the (inverse) Fourier coefficients of the logarithm of reduced {sample spectrum}, $L_0=\langle \log(\xi) \rangle = \psi(\ell-M+1) -\log(\ell-M+1) $, and $\sigma_0^2=\langle \log(\xi)^2 \rangle - L_0^2 = \psi'(\ell-M+1)$, $\psi$ and $\psi'$ being the di- and tri-gamma functions \cite{PolyGamma}, respectively. A thorough derivation of the above formulas as well as a detailed description of the workflow for the analysis of the data produced by EMD simulations are presented in the Supplemental Material (SM) \cite{SupplMat}.}

\begin{figure}
\includegraphics[scale=1.1]{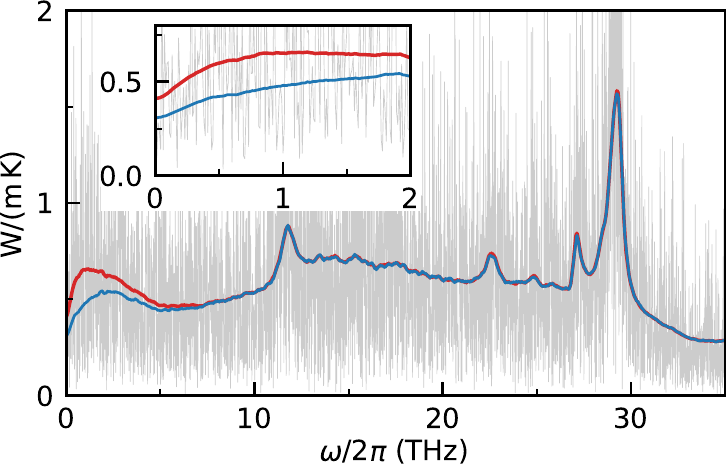}
\caption{Energy flux of a water-ethanol mixture. Gray: energy-flux  {sample} power spectrum from a $100\un{ps}$ trajectory. Red and blue: moving averages over a window of $0.2\un{THz}$ of the energy-flux power  {sample} spectrum  and of the  {SCCB} of the  {sample} cross-spectrum, computed from a long ($28\un{ns}$) trajectory. Inset: low-frequency region of the spectrum.}\label{fig:grappa}
\end{figure}

In order to validate our methodology, we have computed the thermal conductivity of an equimolar water-ethanol mixture. Not aiming at an optimal description of the system, but just at a realistic benchmark, we used the simple OPLSAA \cite{OPLSAA} all-atom flexible force field. Classical EMD simulations were run with the LAMMPS package \cite{PLIMPTON19951} at a temperature of $\approx 350\un{K}$ and a density of $0.80\un{g/cm^3}$, corresponding to $800+800$ molecules in a cubic cell with an edge of $47.47\un{\angstrom}$.

\begin{figure}
\includegraphics[scale=1.1]{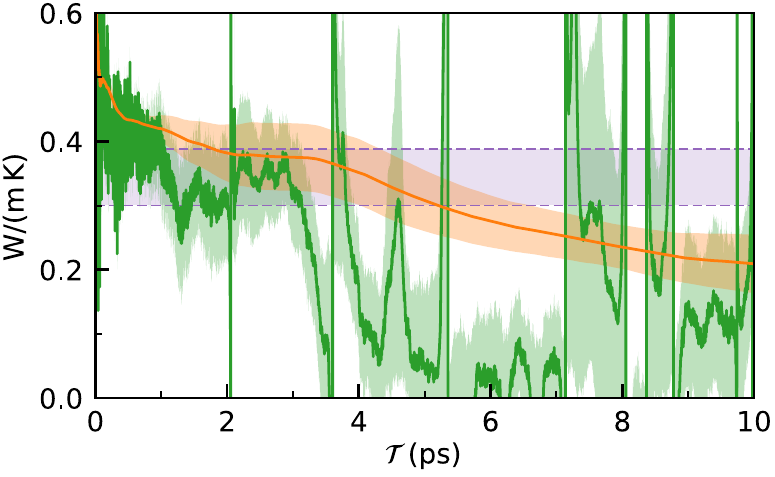}
\caption{Thermal conductivity of a water-ethanol solution, Eq.~\eqref{eq:multi_kappa}, as a function of the upper limit of integration in Eq.~\eqref{eq:L_def}. The shaded areas indicate the estimated statistical error. Green: direct estimate from from Eqs.~\eqref{eq:multi_kappa} and \eqref{eq:L_def}. Orange: estimate from the Einstein-Helfand relation, see main text. Purple: cepstral analysis estimate.} \label{fig:gk-cep}
\end{figure}

In Fig.~\ref{fig:grappa} we report the energy-flux spectrum of our water-ethanol model. The gray line indicates the diagonal element of the raw {sample spectrum}, computed from a $100\un{ps}$ trajectory, which is too noisy to be used as an estimator. Performing a moving average \cite{MovingAverage} of the {sample spectrum} would consistently reduce the noise, at the price however of requiring much longer trajectories (the blue line reports an average performed from a $28\un{ns}$ trajectory) and introducing a bias that is difficult to evaluate and remove. Note that while at high frequency, where the spectrum is dominated by intra-molecular vibrations, the energy-energy diagonal element of the cross-spectrum (red line) and the  {SCCB} coincide, in the low-frequency region, which is characterized by a strongly diffusive behavior, the two differ markedly and only the latter is meaningful to estimate $\kappa$. In Fig.~\ref{fig:gk-cep} we report the thermal conductivity of the  water-ethanol solution computed from  Eq.~\eqref{eq:multi_kappa} as a function of the upper limit of integration in Eq.~\eqref{eq:L_def}, and as obtained from a $100\un{ps}$ EMD trajectory. The spikes in the estimated conductivity result from the vanishing of the $\Lambda^{11}$ denominator in Eq.~\eqref{eq:multi_kappa}, which is in turn due to the fact that the integrals of the correlation functions in Eq.~\eqref{eq:L_def} behave as random walks as soon as the integrand vanishes, eventually assuming any value. This behavior can be partially corrected by replacing the GK estimate of the Onsager coefficients through Eq.~\eqref{eq:L_def} with an equivalent one, based on the Einstein-Helfand relation \cite{Baroni2018,Helfand1960}, which is statistically better behaved (orange line):  {$\Lambda^{ij}=\lim_{\tau\to\infty} \frac{V}{2 k_B\tau}\left \langle \int_0^\tau \mathcal{J}^i(t)dt \; \int_0^\tau \mathcal{J}^j(t)dt \right \rangle$.} Even so, the statistical accuracy that can be achieved with even longer trajectories is totally inadequate.

Cepstral analysis was performed over a $100\un{ps}$ EMD trajectory using the \texttt{ThermoCepstrum} open-source code, which is freely available for download \cite{thermocepstrum}.
We set $\ell=3$ (Cartesian components) and $M=2$ (number of conserved fluxes), thus obtaining $L_0 = \psi(2)-\log(2) \approx -0.270$ and $\sigma_0^2=\psi'(2)\approx 0.644$. The cutoff frequency \cite{Ercole2017} used for cepstral analysis, and ensuring convergence in $\kappa$, was $\omega^*/2\pi \approx 35\un{THz}$. The number of cepstral coefficients  {$P^* \approx 45$ was estimated from the Akaike Information Criterion} \cite{Ercole2017,Akaike1973,*Akaike1974}. The final estimate of the heat conductivity resulting from Eqs.~\eqref{eq:nutshell} is $\kappa=0.34\pm 0.04\un{W/(m\,K)}$. In order to validate our statistical analysis, cepstral analysis was repeated for all the $100\un{ps}$ extracted from a $28\un{ns}$ long trajectory, confirming the normal distribution of the estimated conductivity and the value of the relative error. The small relative error ($13\%$) achieved by analyzing trajectories as short as $100\un{ps}$ shows that cepstral analysis opens the way to heat-transport simulations using \emph{ab initio} EMD even in multi-component systems.
 {Our analysis equally applies to systems with any number of components in any charge state,
because the SCCB is invariant under any nonsingular linear trasformation of the convective fluxes.
This
implies that
in binary molten salts and ionic fluids the electric current, which is a linear combination of mass/number fluxes, can be taken as a proxy of the convective fluxes. An application to molten sodium chloride is presented in the SM \cite{SupplMat}.}

\begin{figure}
\includegraphics[scale=1.1]{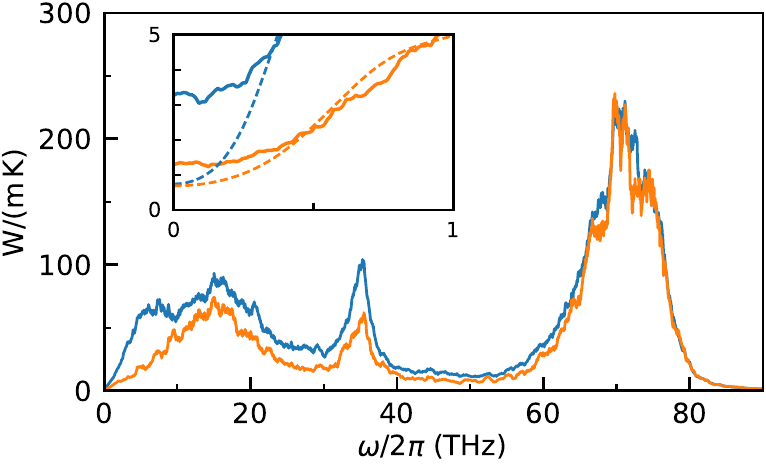}
\caption{Multi-variate analysis of \emph{ab initio} water. The sample power spectrum of the DFT energy flux would be out-of-scale and numerically intractable. Blue:  {SCCB} of a two-component analysis performed with the total momentum of the oxygen atoms as an inert flux. Orange: three-component analysis, where the third flux is the electronic adiabatic current. Both are filtered with a moving average of width $1\un{THz}$. In the inset, the spectrum obtained from cepstral analysis is displayed in dashed lines.} \label{fig:water_multic}
\end{figure}

Multivariate cepstral analysis turns out to be instrumental in heat-transport simulations even for one-component systems. The recently discovered \emph{gauge invariance} of heat conductivity \cite{Marcolongo2016,Ercole2016} means that, while transport coefficients are largely independent of the detailed form of the energy flux, the flux power spectrum and the resulting statistical properties of the estimator do depend on it. The question then naturally arises of how to choose the form of the flux so as to optimize these statistical properties. In the case of \emph{ab initio} water \cite{Marcolongo2016}, for instance, the total power of the energy flux derived from density-functional theory (DFT) is so large as to make the statistical analysis of its spectrum at low-frequency intractable, due to large atomic self-energies \cite{Marcolongo2019} that cannot be easily defined and eliminated in \emph{ab initio} simulations. A couple of \emph{ad hoc} solutions to this problem have been devised \cite{Marcolongo2016,Marcolongo2019}, leveraging gauge invariance to subtract from the energy flux a linear combination of ``inert'' fluxes not contributing to heat transport (such as the electronic flux or the mass flux of one of the two atomic species). While this remedy was effective and actually allowed one to get meaningful results from relatively short trajectories, the question remains of how to optimally choose this linear combination. Multivariate cepstral analysis provides the answer. The idea is to compute, along with the energy flux to be analyzed, a number of inert fluxes and treat all of them on a par as though they were conserved fluxes of a multi-component system: the total power of the  {SCCB} will be considerably reduced, thus making the cepstral analysis feasible at low frequency. This is illustrated in Fig.~\ref{fig:water_multic}, where we report the multi-variate power spectrum of \emph{ab initio} (heavy) water, computed using the total momentum of the oxygen atoms as an inert flux (blue), as well as this and the adiabatic electronic current as inert fluxes (orange). The simulation setting is the same as in Ref.~\citenum{Marcolongo2016}. The spectrum of the bivariate  {SCCB} is always larger than the trivariate one. Using cepstral analysis, the zero-frequency limits of these two spectra coincide, as they must, whereas their running averages do not and depend sensitively on the width of the frequency window. The heat conductivities estimated by cepstral analysis are $(0.74 \pm 0.16)\un{W/(m\,K)}$ and $(0.69 \pm 0.18)\un{W/(m\,K)}$, respectively, while the experimental value is $0.61\un{W/(m\,K)}$ \cite{Matsunaga}.

We conclude by noticing that the combination of the newly devised convective invariance and multivariate cepstral analysis, besides providing fresh theoretical insight in transport phenomena, will hopefully broaden the scope of heat-transport simulations to complex multi-component fluids, as well as provide new tools to make \emph{ab initio} simulations possible.

\medskip
\begin{acknowledgments} This work was partially funded by the EU through the \textsc{MaX} Centre of Excellence for supercomputing applications (Projects No.~676598 and 824143). We are grateful to D. Tisi for a critical reading of the revised version of our manuscript. \end{acknowledgments}

%

\end{document}


\title{Supplemental Material to \\ ``{Theory and numerical simulation of} heat transport in multi-component systems''}

\author{Riccardo Bertossa}
\affiliation{SISSA -- Scuola Internazionale Superiore di Studi Avanzati, Via Bonomea 265, 34136 Trieste, Italy}
\author{Federico Grasselli}
\affiliation{SISSA -- Scuola Internazionale Superiore di Studi Avanzati, Via Bonomea 265, 34136 Trieste, Italy}
\author{Loris Ercole}
\affiliation{SISSA -- Scuola Internazionale Superiore di Studi Avanzati, Via Bonomea 265, 34136 Trieste, Italy}
\author{Stefano Baroni}\email{baroni@sissa.it}
\affiliation{SISSA -- Scuola Internazionale Superiore di Studi Avanzati, Via Bonomea 265, 34136 Trieste, Italy}
\affiliation{CNR -- Istituto Officina dei Materiali, SISSA, 34136 Trieste}

\date{\today}

\maketitle

\section{S1 -- Multivariate cepstral analysis}

Practical numerical equilibrium molecular dynamics (EMD) simulations provide access to \textit{discrete} samples of several flux process, \textit{i.e.}~to finite multivariate \textit{time series}. 
Let us then suppose to have $\ell$ independent such samples, of length $N$: $\{ ^{p\!}\mathcal{J}^i_n \} ~(p=1,\dots \ell; ~i=0,\dots M-1; ~n=0,\dots N-1)$. For instance, the three Cartesian components of a current in an isotropic medium may be thought of as different samples of a same process, and so would different segments of a same long time series. Stationarity implies that $\langle {\mathcal J}^i_n\rangle $ does not depend on $n$ and that $\langle {\mathcal J}^i_n {\mathcal J}^j_m \rangle$ only depends on $n-m$. We further assume that $\langle {\mathcal J}^i_n\rangle =0 $ and that $\langle {\mathcal J}^i_n {\mathcal J}^j_0 \rangle$ is an even function of $n$, which is the case when ${\mathcal J}^i$ and ${\mathcal J}^j$ have the same signature under time-reversal. We define the \emph{cross-periodogram} as the average  discrete-frequency spectrum of the $\ell$ samples:
\begin{equation}
\mathcal{S}_k^{ij}= \frac{1}{\ell} \sum_{p=1}^{\ell} \frac{\epsilon}{N} \left({}^{p\!}\tilde{\mathcal{J}}_k^i\right)^* {}^{p\!}\tilde{\mathcal{J}}_k^j, \label{eq:X-periodogram}
\end{equation}
where 
\begin{equation}
    ^{p\!}{\tilde{\mathcal{J}}}_k^i = \sum_{n=0}^{N-1} {}^{p\!}\mathcal{J}^i_n \mathrm{e}^{2\pi i\frac{kn}{N}}
\end{equation}
is the discrete Fourier transform of the $p$-th time series and $\epsilon$ its sampling time. 
The Wiener-Khintchine theorem \cite{Wiener1930} states that $\mathcal{S}_k^{ij}$ is an (asymptotically) unbiased estimator of the cross-spectrum $S^{ij}(\omega) = \int_{-\infty}^{+\infty} \mathrm{e}^{i\omega t} \langle \mathcal{J}^i(t) \mathcal{J}^j(0) \rangle dt$, \textit{i.e.}~ $S^{ij} = \lim_{N\to \infty} \langle \mathcal{S}^{ij} \rangle$. 
In the large-$N$ limit the real and imaginary parts of $\tilde{\mathcal{J}}^i_k$ are normal deviates that are uncorrelated amongst themselves and for $k\ne k'$. Therefore, for each discrete frequency $k$, the cross-periodogram is a random matrix distributed as a complex Wishart deviate \cite{Goodman1963a,*Goodman1963b,*Wishartdef}: $ \mathcal{S}_k \sim \mathcal{CW}_M\bigl (S(\omega_k), \ell\bigr )$, where, in the literature on time-series analysis, the variance matrix $S(\omega_k)$ is called the cross power spectrum. 
We have the following notable property of Wishart deviates: \textit{the Schur complement of each of the diagonal blocks of a Wishart deviate is also a Wishart deviate} \cite{*[{This property is demonstrated in the real case, \emph{e.g.,} in }] [{, Propositions 7.5 p. 90, and 7.9, p. 92. The generalization to the complex case is trivial.}] bilodeau1999}. More explicitly, let $\mathcal{S} \sim \mathcal{CW}_M\bigl (S, \ell\bigr )$ be a Wishart matrix of rank $M$ and $\bar{\mathcal{S}}$ the Schur complement of rank $L$ of any of its diagonal blocks of rank $M-L$;
one then has:
\begin{equation}
    \bar{\mathcal{S}} \sim \mathcal{CW}_L\bigl (\bar{S}, \ell-M+L\bigr ).
\end{equation}
The Schur complement of the convective-current block (SCCB) of the cross-periodogram, $\bar{\mathcal{S}}^0$, (dubbed the \textit{reduced periodogram}) has rank $L=1$. Therefore, since the Wishart distribution for rank-one matrices is a $\chi^2$ distribution, we conclude that, for each frequency $k$,
\begin{equation}
  \bar{\mathcal S}_k^{0} \sim \bar{S}^{0}_k \, \xi_k,  \label{eq:Supp:mean-multi-periodogram}
\end{equation}
where the $\xi$'s are independent identically distributed (iid) random variables, $\xi \sim \frac{1}{\nu} \chi^2_{\nu}$, $\chi^2_\nu$ being the chi-square distribution with $\nu=2(\ell-M+1)$ degrees of freedom. Strictly speaking, this distribution holds for $k\notin \{0,\frac{N}{2}\}$ because for these two frequencies the Fourier components of the flux time series are real, and $\nu=\ell-M+1$. For the purposes of the following discussion, we overlook this exception, thus making an overall error $\sim{\mathcal O}(1/N)$, which vanishes  in the $N\to\infty$ limit that is assumed throughout our statistical analysis. Moreover for clarity we included a factor $\ell / \nu$ in the SCCB, that is needed to obtain the correct mean value, see Ref.~\citenum{Baroni2018}. 

Equation \eqref{eq:Supp:mean-multi-periodogram} shows that the SCCB of the cross-periodogram (that we dub \textit{reduced} periodogram) is an unbiased estimator of the SCCB of the cross-spectrum (the \textit{reduced} spectrum), which is proportional to $\kappa$. Unfortunately, this estimator is not consistent, in that its variance does not vanish as $N\to\infty$ (actually, it is independent of $N$).
A consistent estimator is obtained by applying a low-pass filter to the \textit{logarithm} of the reduced periodogram, thus introducing a bias, which however can be made arbitrarily small as $N\to \infty$.
Eq.~\eqref{eq:Supp:mean-multi-periodogram} shows that 
\begin{equation}
    \log\bigl ( \bar{\mathcal S}_k^{0} \bigr ) -\log \bigl ( \bar{S}^{0}_k \bigr )\sim \log(\xi)
\end{equation} 
are iid stochastic variables whose expectation and variance are: $L_0=\langle \log(\xi) \rangle = \psi(\ell-M+1) -\log(\ell-M+1) $ and $\sigma_0^2=\langle \log(\xi)^2 \rangle - L_0^2 = \psi'(\ell-M+1)$, $\psi$ and $\psi'$ being the di- and tri-gamma functions \cite{PolyGamma}, respectively. In order to implement such a filter, we define the energy-flux \emph{cepstrum} as the inverse Fourier transform of $\log \bar{\mathcal S}^{0}$:
\begin{equation}
  \mathcal{C}_n =\frac{1}{N} \sum_{k=0}^{N-1}\log \bigl (
  \bar{\mathcal{S}}^0_k \bigr ) \mathrm{e}^{ - 2\pi i\frac{kn}{N}}. \label{eq:cepstral-multi-def}
\end{equation}
According to a generalized central-limit theorem for the Fourier transform of iid stochastic variables \cite{Anderson1994,*Peligrad2010}, for $N\to\infty$ the \emph{cepstral coefficients}, $\mathcal{C}_n$, are iid normal variates whose expectation value is 
\begin{equation*}
    \begin{split}
        \langle\mathcal{C}_n\rangle&=\frac{1}{N} \sum_{k=0}^{N-1}\left[ \langle\log(\xi)\rangle +  \langle\log\bigl ( \bar{S}_k^{0} \bigr )\rangle \right] \mathrm{e}^{ - 2\pi i\frac{kn}{N}} \\
        &=\frac{1}{N} \sum_{k=0}^{N-1}\left[ L_0 +  \langle\log\bigl ( \bar{S}_k^{0} \bigr )\rangle \right] \mathrm{e}^{ - 2\pi i\frac{kn}{N}}\\
        &=L_0\delta_{n0} + C_n ,
    \end{split}
\end{equation*}
where $C_n=\frac{1}{N} \sum_{k=0}^{N-1}\log \bigl ( \bar{S}^0_k \bigr ), \mathrm{e}^{-2\pi i\frac{kn}{N}}$, and the variance, that can be computed in a similar way, is $ \langle\mathcal{C}_n^2 \rangle - \langle\mathcal{C}_n \rangle^2=\frac{1}{N}\sigma_0^2$.
The assumption that $S$ is a smooth function implies that the number of significant cepstral coefficients $\mathcal{C}_n$ in the cepstral expansion of Eq.~\eqref{eq:cepstral-multi-def} is hopefully \textit{small}, allowing us to implement here a low-pass filter. 
We thus indicate by $P^*$ a small integer such that
\begin{equation}
C_n\approx 0 \;\;\;\;\text{  for  }\;\;\;\; P^*\le n\le N-P^*.
\label{eq:filter}
\end{equation}
Unfortunately, the expectation value of the inverse Fourier transform of $\mathcal C_n$ does not represent $\log(\bar{S}^0_k)$, because of the additional $L_0$ factor:
\begin{equation*}
    \left\langle \sum_{n=0}^{N-1} \mathcal C_n \mathrm{e}^{ 2\pi i\frac{kn}{N}} \right\rangle = L_0 + \sum_{n=0}^{N-1} C_n \mathrm{e}^{ 2\pi i\frac{kn}{N}}.
\end{equation*}
Since we are interested in the $k=0$ component and Eq.~\eqref{eq:filter}, we must consider the stochastic variable
\begin{equation}
\mathcal{L}^0 = -L_0 + \mathcal{C}_0 + 2 \sum_{n=1}^{P^*-1} \mathcal{C}_n,
\end{equation}
whose expectation value and variance are
\begin{equation}
\begin{split}
    \langle \mathcal{L}^0 \rangle &= \log \left ( \bar{S}^0_0\right ) \\ 
    \mathrm{var}(\mathcal{L}^0) &=\frac{4P^*-2}{N}\sigma_0^2.
\end{split}
\end{equation}
$\mathcal{L}^0$ is thus an \textit{unbiased estimator} for the logarithm of the SCCB in the zero-frequency cross-spectrum of the conserved fluxes and, hence, of the thermal conductivity we are after. At fixed $P^*$, the variance of $\mathcal{L}^0$ vanishes in the $N\to\infty$ limit, thus making the estimator consistent. Of course, any finite choice of $P^*$ introduces a bias, which is a decreasing function of $P^*$, while the statistical error is an increasing function of it. Its optimal value is the one which makes the bias of the order of the statistical error. By adopting this value in the $N\to\infty$ limit both the bias and the statistical error can be made arbitrarily small. 

An estimate of the ``optimal'' value for $P^*$ can be obtained with any of the many \emph{model selection} techniques available in the statistics literature \cite{*[{For a review of model selection techniques, see \emph{e.g.}}] [], Claeskens2008}; in our applications we choose to adopt the Akaike information criterion (AIC) \cite{Akaike1973,*Akaike1974,Ercole2017}, even if other more sophisticated, and possibly more efficient, methods could be devised.

\section{S2 -- Application to a molten salt}

In this section we present the results obtained for molten sodium chloride (NaCl), paradigmatic of a two-component (neutral) mixture of charged particles.
As for the water-ethanol mixture, classical MD simulations were run with the LAMMPS package \cite{PLIMPTON19951} in the NVE ensemble. For NaCl we used the Fumi-Tosi \cite{FumiTosi-1,*FumiTosi-2} classical force field, and ran simulations at a temperature of $\approx 1400\un{K}$ and a density of $1.29\un{g/cm^3}$, corresponding to $864$ formula units in a cubic cell with an edge of $40.21\un{\angstrom}$, and a simulation time step of $1\un{fs}$.
The cepstral analysis was again performed over blocks of $100\un{ps}$. 
The energy flux is computed via the standard expression 
adopted in the LAMMPS package.
A $34\un{ns}$ long trajectory was employed to extract many $100\un{ps}$ blocks, in order investigate the statistics of the resulting estimates for the transport coefficients.
\begin{figure}
\begin{center}
    \includegraphics[width=\linewidth]{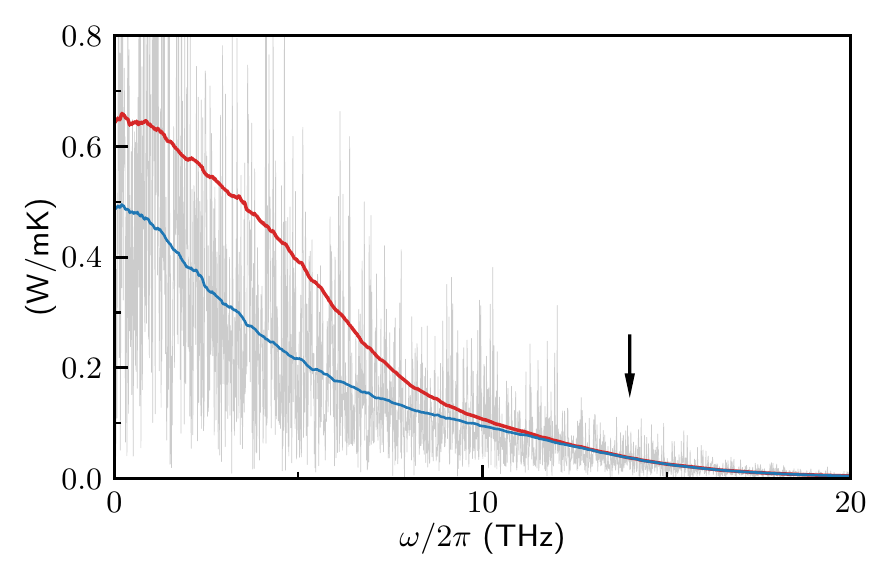}
\caption{
Energy flux of molten NaCl. Gray: heat flux periodogram from a $100\un{ps}$ trajectory segment. Red and blue: moving average over a window of $0.2\un{THz}$ of the energy-flux periodogram, and moving average of the reduced periodogram $\bar{\mathcal{S}}^0(\omega_k)$, respectively, computed over the whole trajectory ($34\un{ns}$).
The vertical arrow indicates the value of the Nyqvist frequency used to perform the cepstral analysis.
}\label{fig:S1psd}
\end{center}
\end{figure}
In Fig.~\ref{fig:S1psd} we show the (moving average of the) energy flux periodogram, $\mathcal{S}^0_k$ (red), along with the reduced periodogram $\bar{\mathcal{S}}^0_{k}$ (blue), whose expectation value is proportional to the thermal conductivity in conditions of zero charge current (see Eq.~(3) in the main manuscript), according to 
\begin{equation}
    \kappa = \frac{V}{2k_B T^2} \bar{S}^0(\omega = 0). \label{eq:Supp-multi_kappa}
\end{equation}
We notice that the red and blue curves substantially differ at low frequencies, thus giving a different zero-frequency value, and demonstrating the importance of employing Eq.~\eqref{eq:Supp-multi_kappa} in place of $\kappa = \frac{V}{2k_B T^2} S^{00}(0)$, when there is more than one conserved flux.
From a standard Green-Kubo analysis of the whole trajectories ($34\un{ns}$) we obtained $\kappa_{\mathrm{ref}} = (0.485\pm 0.005)\un{W/mK}$, that can be considered as a reference.
Furthermore, for each of the $100\un{ps}$ segments we computed $\bar{\mathcal{S}}^0_k$, and obtained an estimate $\bar{S}^0(0)$ and the thermal conductivity $\kappa$, according to Eq.~\eqref{eq:Supp-multi_kappa}.
We set $\ell=3$ (Cartesian components) and $M=2$ (number of conserved fluxes) to define the theoretical statistical distributions necessary for the cepstral analysis. As in the case of the water-ethanol mixture, the parameters are $L_0 = 1-\gamma-\log 2 \approx 0.270$ and $\sigma_0^2=\pi^2/6-1\approx 0.644$.
\begin{figure}
\begin{center}
    \includegraphics[width=\linewidth]{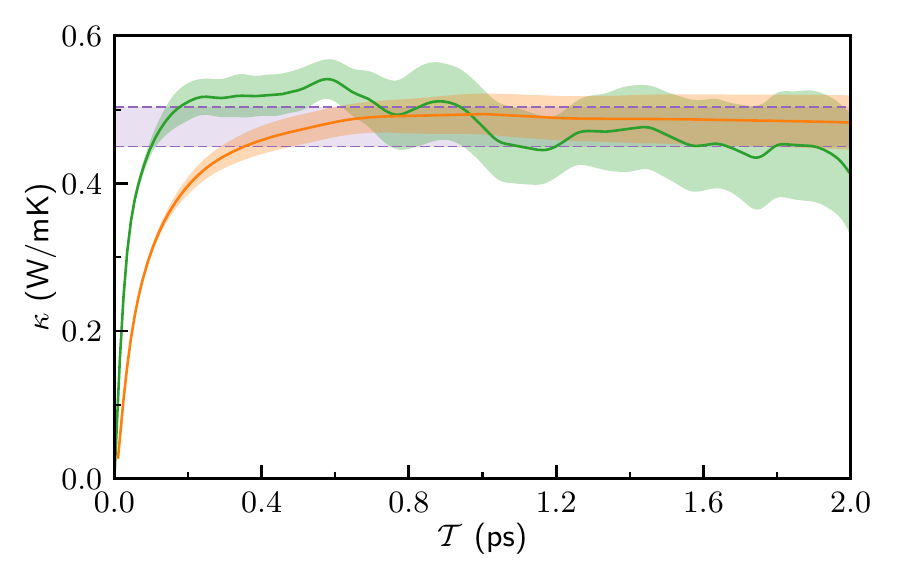}
\end{center}
\caption{
Thermal conductivity estimator for molten NaCl, as a function of the upper limit of time-integration. The shaded ares surrounding the lines indicate one standard deviation, as estimated by a standard block analysis over all the trajectory segments.
Green: direct time-integration according to Eqs.~(2-3) in the main text.
Orange: direct time-integration of the Einstein-Helfand integrals.
Purple: cepstral analysis estimate from the reduced periodogram.
	\label{fig:S2gk} }
\end{figure}
For the cepstral analysis we employed the cutoff frequency $f^*\approx 14\un{THz}$. The estimate of $\kappa$ has been verified to not depend on the specific value of $f^*$. The number of cepstral coefficients $P^*$ has been estimated by the optimization of the AIC \cite{Ercole2017,Akaike1973,Akaike1974} for each segment independently, and the average number of coefficients is: $P_A^* = 5$.
Finally, the relative error on $\kappa$ estimated by the second of Eqs.~\eqref{eq:nutshell} is $6.5\%$.
The average value of thermal conductivity of NaCl, computed over all the blocks from cepstral analysis, is $(0.487 \pm 0.002)\un{W/mK}$, in perfect agreement with the reference values obtained above.

\section{S3 -- Workflow}

In a nutshell, our cepstral method requires the following list of operations, summarized in this Section for clarity:
\begin{enumerate}
    \item Extract $\ell$ independent samples of $N$ (equally spaced) steps of the time series for the energy flux $\mathcal{J}^0_n$, $n=1,\ldots,N$, and for a complete set of independent particle fluxes $\mathcal{J}^m_n$, $m=1,\ldots,M-1$.
    \item Compute the discrete Fourier transforms of these fluxes, $\tilde{\mathcal{J}}^i(\omega_k)$, the  cross-periodogram $\mathcal{S}^{ij}(\omega_k)$, $i,j=0,\ldots,M-1$, and the SCCB of the cross-periodogram (with the factor $\ell / \nu$) for each frequency, $\bar{\mathcal{S}}^0(\omega_k)$, where $k=1,\ldots,N$.
    \item Calculate the logarithm of the SCCB of the cross-periodogram, $\log \bar{\mathcal{S}}^0(\omega_k)$.
    \item Compute the inverse discrete Fourier transform of the result to obtain the cepstral coefficients $\mathcal{C}_n$.
    \item Retain the first $P^*$ cepstral coefficients, the optimal number $P^*$ being computed via the Akaike information criterion \cite{Ercole2017,Baroni2018}.
    \item Evaluate the thermal conductivity and its statistical uncertainty estimated as:
\begin{equation}
    \begin{gathered}
       \mathcal{K} = \frac{V}{2k_BT^2}\mathrm{exp}\left [ \mathcal{C}_0 + 2\sum_{n=1}^{P^*-1}\mathcal{C}_{n} -L_0\right ] \\
        \frac{\Delta\kappa}{\kappa} = \sigma_0 \sqrt{\frac{4P^*-2}{N}},
    \end{gathered} \label{eq:nutshell}
\end{equation}
where $L_0$ and $\sigma_0$ are properties of the $\chi^2$ distribution, as discussed before.  
\end{enumerate}
The relative error in the conductivity results from the estimate of the absolute error in its logarithm. $P^*$ depends in general on the cutoff (Nyqvist) frequency, $f^\ast$, used to analyze the spectrum, while the final value of $\kappa$ and its statistical uncertainty are largely independent of it \cite{Ercole2017,Baroni2018}.

This workflow of operations is implemented in the \texttt{ThermoCepstrum} code \cite{thermocepstrum}, for which we provide an example folder at \url{https://github.com/lorisercole/thermocepstrum}.

\bigskip


%